\documentclass[aps,pra,twocolumn,amsmath,amssymb,letterpaper,groupaddresses,superscriptaddress]{revtex4}
\usepackage{times}
\usepackage{latexsym}
\usepackage{graphicx}
\usepackage{verbatim,times,bbm}
\usepackage{color}
\usepackage{appendix}
\usepackage{bm}
\usepackage{bbm}
\usepackage{amsmath}

\def\uno{\mbox{1 \kern-.59em {\rm l}}}

\def\be{\begin{equation}}
\def\ee{\end{equation}}
\def\ba{\begin{eqnarray}}
\def\ea{\end{eqnarray}}

\def\h{\hskip 1cm }

\def\la{\langle}
\def\ra{\rangle}
\def\a{\alpha}
\def\b{\beta}

\def\l{\lambda}

\def\h{\hat}

\begin{document}

\title{Sequentially generated entanglement, macroscopicity and squeezing in a spin chain}

\author{Tahereh Abad}
\affiliation{Department of Physics, Sharif University of Technology, Tehran, Iran}
\affiliation{Department of Physics and Astronomy, Aarhus University, 8000 Aarhus C, Denmark}
\author{Klaus M\o lmer}
\affiliation{Department of Physics and Astronomy, Aarhus University, 8000 Aarhus C, Denmark}
\author{Vahid Karimipour}
\affiliation{Department of Physics, Sharif University of Technology, Tehran, Iran}

\begin{abstract}
We study quantum states generated by a sequence of nearest neighbor bipartite entangling operations along a one-dimensional chain of spin qubits. After a single sweep of such a set of operations, the system is effectively described by a matrix product state (MPS) with the same virtual dimension as the spin qubits. We employ the explicit form of the MPS to calculate expectation values and two-site correlation functions of local observables, and we use the results to study fluctuations of collective observables. Through the so-called macroscopicity and the squeezing properties of the collective spin variables they witness the quantum correlations and multi-particle entanglement within the chain. Macroscopicity only occurs over the entire chain if the nearest neighbor interaction is maximally entangling, while a finite, sequential interaction between nearest neighbor particles leads to squeezing of the collective spin.  \end{abstract}

\maketitle 

\section{Introduction}\label{intro}
To emphasize what they saw as serious problems within the quantum formalism, Erwin Schr\"odinger and Albert Einstein presented "paradoxical" situations, such as a cat being simultaneously dead and alive if subject to a poison administered by an atomic trigger mechanism \cite{Schrodinger1}, and particles whose state would be magically "steered" by an experimentalist acting on another, remote and physically detached particle \cite{Schrodinger2, EPR, Schrodinger3,  Wiseman}. The phenomena sketched by these situations played a significant role and shaped the way that we interpret and discuss quantum phenomena until this very day. While being originally confined to Gedankenexperiments and discussions on interpretation, they have also become the basis for candidate quantum technology applications. Quantum systems occupying macroscopically separated state components thus hold potential for high precision sensing, while two- and many-particle entangled states have applications for quantum communication and information processing. With these applications comes also the need to quantify physical properties, and in this article we shall consider two of these properties, namely macroscopicity and squeezing, both of which are related to fluctuations of additive observables. While macroscopicity deals with large fluctuations obtained for superpositions or mixtures of macroscopically distinct states \cite{Dur, Shimizu and Miyadera, Korsbakken, Marquardt, Lee}, multi-partite entangled states with a certain minimum number of individual systems, a certain depth of entanglement, may be witnessed by the reduced, squeezed, fluctuations of collective spin observables \cite{Sorensen,Toth}.\\

We deal with the fluctuations of an additive observable $A=\sum_{i}A_{i}$, where $A_i$ denote observables acting on $N$ different subsystems. For an uncorrelated state we have the variance ${\cal V}(\sum_{i}A_{i})=\sum_{i}{\cal V}(A_{i})$, which scales linearly with $N$, while correlations among the subsystems may yield scaling with a higher or lower power of $N$, ${\cal V}(\sum_{i}A_{i})\propto N^\alpha$. An asymptotic scaling with $N^\alpha$, with $\alpha \ne 1$, is possible if a fraction of the particles are correlated \cite{Frowis, Shimizu} and occurs, for example, at the critical point of a system undergoing a phase transition \cite{abad}. Fluctuations scaling as $N^2$ may be due to classical correlations, e.g., of particles that all occupy either one or another state, while if such scaling is observed in a pure quantum state, the systems must be entangled.
Sets of inequalities have been derived that must be obeyed by the fluctuations of one or several collective observables of non-entangled quantum systems, see, e.g, \cite{Sorensen,Toth}. In particular squeezing, i.e, the reduced fluctuations of collective observables compared to the independent particle case, has been promoted as a criterion for entanglement between particles, and witnessing their potential use in interferometric applications \cite{Bjork}.
\\
\begin{figure}[h]\label{chain}
\centering
\includegraphics[width=5.5cm,height=3.5cm,angle=0]{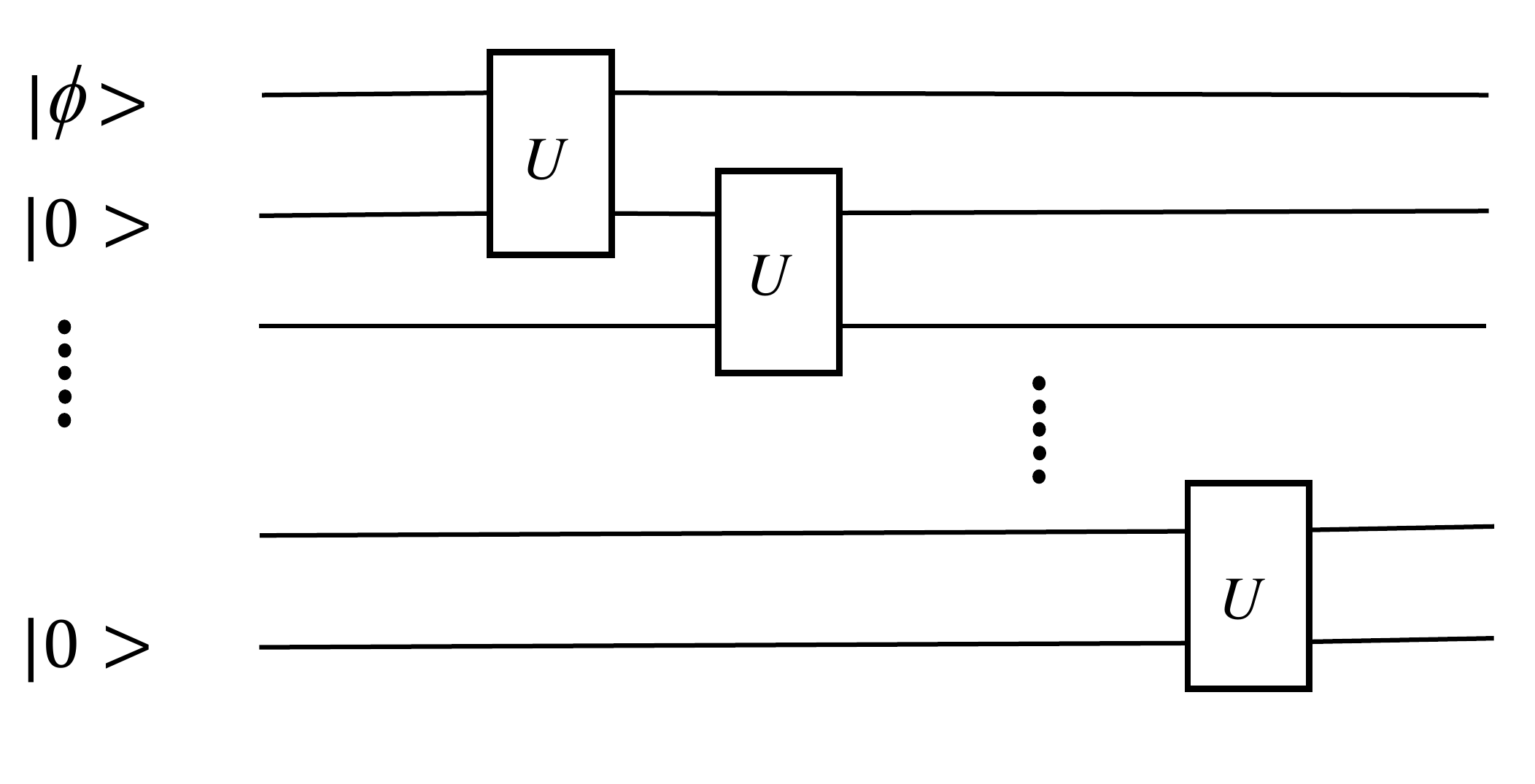}
\caption{A two-qubit operation $U$ is sequentially applied between the nearest neighbor qubits along a one dimensional chain. We assume that the first qubit initally occupies a state $|\phi\ra=c_0|0\rangle+c_1|1\rangle$, while all the other qubits occupy the state $|0\rangle$.}
\end{figure}

In this article we consider a special class of states, created by the sequential application of operations acting on pairs of initially uncorrelated spins or qubits along a one dimensional chain, see Fig.1. This class of states can also be prepared by sequential interaction of the qubits with a single travelling qubit, or by the emission of qubits from a common origin \cite{Cirac1}. Although the interaction Hamiltonian only addresses each pair of particles once, the system may become entangled over long distances and hence develop a rather complex character. However, states of precisely this kind are particularly suitable to be represented by a Matrix Product State (MPS) \cite{Cirac1, Cirac2, Cirac3} and we provide the explicit MPS, for any given two-qubit unitary applied along the chain. The MPS representation directly yields one- and two-site expectation values and lead to the mean and variance of additive observables. We characterize which two-qubit unitaries lead to macroscopicity, and how much squeezing and entanglement results from a particularly chosen interaction Hamiltonian.

The article is organized as follows. In Sec. \ref{Sec2}, we discuss measures and examples of macroscopicity and the relation between spin squeezing inequalities and entanglement. In Sec. \ref{Sec3}, we present the MPS produced by our sequential preparation procedure. In Sec. \ref{Sec4}, we determine mean values and correlation functions from the MPS description, and we quantify the degree of macroscopicity. In Sec. \ref{Sec5}, we discuss spin squeezing and we show that global observables of the system suffice to witness that the pairwise entangling operations can produce states with an entanglement depth larger than two, i.e., the system contains more than pairwise entanglement. We conclude with a brief outlook in Sec. \ref{Sec6}.
\section{Quantum macroscopicity, squeezing and entanglement}\label{Sec2}

In this section we briefly review the basic notions of macroscopicity and squeezing. Although there are various measures for quantifying these properties, we focus on two specific measures \cite{Frowis, Sorensen} which are most suitable for our system of interest, which is a collection of $N$ spin $1/2$-particles.

\subsection{Measure of macroscopic quantum superposition}\label{SubSec1}
The property of macroscopicity \cite{Shimizu} is most pronounced in a Greenberger-Horne-Zeilinger ($GHZ$) state of $N$ qubits,
\begin{equation}
|GHZ\rangle=\frac{1}{\sqrt{2}}(|0\rangle^{\otimes N}+|1\rangle^{\otimes N}),
\end{equation}
which is a superposition of two macroscopically distinct state. This state
features an extensive variable $Z=\sum^N_{i=1} \sigma^z_i$ with an anomalously large variance: $(\bigtriangleup Z )^{2}=\langle Z^2\rangle-\langle Z \rangle^2=N^2$, where $\sigma^{z}_{i}$ is the Pauli operator acting on site $i$. This is to be compared with states of the form $[\frac{1}{\sqrt{2}}(|0\rangle+|1\rangle)]^{\otimes N}$, where superposition (in the same basis) exists only on a small scale (on the level of individual spins). For such states and similar ones (where a few spins are in superposition) the variance scales only linearly with the system size $N$. Of course the fluctuations depend on which observable is considered, and for a more general definition one considers the maximally fluctuating quantity and introduce the concept of an "effective size" of a state $|\psi \rangle$ of $N$ particles, \cite{Frowis}
 \begin{equation}\label{effSize}
N_{\text{eff}}(\psi)=\max_{A} (\bigtriangleup A )^{2}/N.
 \end{equation}
where $(\bigtriangleup A )^{2}=\langle \psi| A^2 |\psi \rangle-\langle \psi| A |\psi \rangle^2$, and the variance of $A$ is maximized over all sums of local operators $A=\sum^N_{i=1} A_i$, where $A_i$ acts on the $i$th particle and has eigenvalues $\pm 1$.

$N_{\text{eff}}$ defines the scale over which macroscopic superpositions and hence quantum behavior prevail. If $N_{\text{eff}}={\cal O}(N)$ , we have a macroscopically correlated state while, if $N_{\text{eff}}={\cal O}(1)$, quantum behavior only manifests itself at the microscopic level of few particles.

Note that the definition of macroscopicity involves a maximization over possibly different operators acting on the different particles. This makes it generally difficult to calculate and practical settings, e.g., symmetrical addressing by the preparation and measurement procedures, may justify the simplifying assumption of identical $A_i$ operators.


\subsection{Spin squeezing inequalities and entanglement} \label{SubSec2}

Rather than increased fluctuations, also reduced fluctuations reveal correlations within ensembles of quantum particles. Such squeezing holds potential for applications in high precision measurements and quantum information processing, and a host of methods exist to entangle and squeeze collective spin degrees of freedom of atomic, electronic or nuclear ensembles, e.g., to reduce spectroscopic noise or to improve the accuracy of atomic clocks \cite{Wineland}.

The basic concept of squeezed spin states was established in \cite{Kitagawa}. If the mean collective spin vector is aligned with the $z$ axis, the variances of the orthogonal components are bounded by the uncertainty relation
\begin{equation}
(\bigtriangleup J_{x})^{2}(\bigtriangleup J_{y})^{2}\geq \frac{1}{4}|\langle J_{z}\rangle|^{2},
\end{equation}
where $\hbar=1$ and $J_{\mu}=\frac{1}{2}\sum_{i=1}^{N}\sigma_{i}^{\mu}$, where $\sigma_{i}^{\mu}$ is a Pauli matrix ($\mu=x,y,z$) at site $i$. If, for example,  $(\bigtriangleup J_{x})^{2}$ is smaller than the standard quantum limit $\frac{1}{2}|\langle J_{z}\rangle|$, the state is spin squeezed.

The relevance of comparing the length of the spin with the fluctuations of a perpendicular component is clear in Ramsey spectroscopy, where the mean signal oscillates with an amplitude proportional to the initial length of the spin, say $|\langle J_z \rangle |$, and shows the largest variation when the projection measured vanishes. At this point the signal fluctuations  are governed by the variance of the orthogonal spin components. Wineland \textit{et al.} \cite{Wineland} have shown that the measurement resolution in atomic clocks depends on the spin squeezing parameter
\begin{equation}\label{xi}
\xi^{2}=\frac{N (\bigtriangleup J_{\theta})^{2}}{\langle J_{z}\rangle^{2}},
\end{equation}
where $J_{\theta}=\cos(\theta) J_{x}+\sin (\theta) J_{y}$ denotes the experimentally relevant spin component in the plane orthogonal to the mean spin.

The inequality $\xi^{2} < 1$ indicates that the system is spin squeezed, and it has been shown that any state with this property is an entangled state \cite{Duan1,Duan2}. More detailed studies \cite{Sorensen} have shown that the pair of values $(\bigtriangleup J_{\theta})^{2}$ and $|\langle J_{z}\rangle|$ can be used to quantify the entanglement depth, i.e., how large sub-ensembles of spins must at least be entangled to account for the macroscopic mean values and fluctuations.
These and related criteria \cite{Toth} have led to the experimental demonstration of entanglement encompassing hundreds and thousands of particles, \cite{Gross, Lucke, Schmied, Parisa}. We shall show that the sequential interactions between pairs of spin qubits along the chain in Fig.1, also leads to spin squeezing and entanglement, verifiable by the global mean and variances.
\section{Sequential generation of a state and its matrix product form}\label{Sec3}
Consider a chain of spins in an initial pure product state, see Fig.1, on which we sequentially act by a local unitary operator. That is, we perform first a unitary operation affecting the sites $1$ and $2$, then $2$ and $3$, etc., until $N-1$ and $N$ (the same class of states can be generated by a controllable ancillary particle that interacts sequentially with all the spins).

Here we will show that the resulting state can be cast in a matrix product form with very special properties.
We describe our system as qubits with basis vectors $\{|0\rangle,|1\rangle\}$, and with the chain of $N$ qubits prepared in the product state $|\psi_{0}\rangle=|\phi\ra\otimes|0\rangle^{N-1}$, where the first qubit occupies a superposition state $|\phi\ra=\sum_{i=0,1}c_i|i\ra$. Performing the two-qubit unitary $U=\sum_{ijkl}U_{ij,kl}|ij \rangle \langle kl|$ on the first two qubits in the chain results in
\be
U_{12}|\phi\ra|0,0\cdots \rangle=\sum_i c_iU_{m,n;i,0}|m,n,0,0,\cdots\rangle,
\ee
where $U_{12}$ signifies that the unitary $U$ acts on the qubits $1$ and $2$. Acting by $U_{23}$, we find
\be \label{U321}
U_{23}U_{12}|\phi\ra|0,0\cdots \rangle=\sum_{\substack{i,m,n\\  p,q}} c_iU_{m,n;i,0}U_{p,q;n,0}|m,p,q,0,\cdots\rangle.
\ee              
Defining the matrices
\be \label{DefV}
(V_{i})_{jk}:=U_{ik,j0},
\ee
Eq. (\ref{U321}) can be written in the form
\be
U_{23}U_{12}|\phi\ra|0,0\cdots \rangle=\sum_{\substack{i,m\\  p,q}}c_i(V_mV_p)_{i,q}|m,p,q,0,\cdots\rangle.
\ee
and the state after sequential action of $N-1$ operators can be written as a matrix product state \cite{MPS},
\ba \label{psii}
|\psi\rangle&=&U_{N-1,N}...U_{12}|\psi_{0}\rangle\\ \nonumber
&=&\sum_{i,\{i_j\}_{j=1}^N} c_i(V_{i_{1}}...V_{i_{N-1}})_{i, i_N}|i_{1},i_{2},...,i_{N}\rangle,
\ea
or equivalently to
\be \label{psii}
|\psi\rangle =\sum_{\{i_j\}_{j=1}^N} tr(V_{i_{1}}...V_{i_{N-1}}W_{i_{N}})|i_{1},...,i_{N}\rangle,
\ee
where
\be \label{DefW}
W_{i_N}=|i_N \rangle \langle \phi^*|,
\ee
where $\langle \phi^*|=\sum_i c_{i} \langle i|$. Note that the unitarity of $U$,  leads to the constraint
\be \label{constrain}
V_{0}^* V_{0}^T+ V_{1}^*V_{1}^T=1.
\ee
and to the unitality of the completely positive map
\be \label{unitalMap}
{\cal E}(\sigma):=\sum_{i=0}^1 V_i^*\sigma V_i^{T},
\ee
and to the trace-preservation of its dual map
\be \label{dualmap}
{\cal E^*}(\rho):=\sum_{i=0}^1 V_i^{T}\rho V_i^*.
\ee
Let us denote the vectorized form of any matrix $\sigma=\sum_{i,j}\sigma_{i,j}|i\ra\la j|$ by $|\sigma\ra:=\sum_{i,j}\sigma_{i,j}|i, j\rangle$, then it is readily found that
\be
|{\cal E}(\sigma)\ra=E|\sigma\ra,
\ee
where
\begin{equation}\label{DefE1}
E= \sum_{i=0,1} V_{i}^{*} \otimes V_{i}.
\end{equation}
The matrix representation of $E$ allows efficient calculation of its spectrum and of its subsequent action along the chain of qubits. From (\ref{constrain}), we note that the identity matrix $I$ is mapped into itself by (\ref{unitalMap}), and hence its vectorized form
\be
|I\ra=|00\ra+|11\ra,
\ee
 is an eigenvector of $E$ with eigenvalue equal to $1$
\be
E|I\ra=|I\ra.
\ee
From the isospectrality of the dual maps ${\cal E}$ and ${\cal E}^*$ and trace-preserving property of the latter, we find that all the eigenvalues of ${\cal E}$ and hence $E$ must have modulus less than or equal to one.

 It is useful to introduce 
 \be \label{XMatrix}
 X=\sum_{i=0,1} W_{i}^{*} \otimes W_{i},
 \ee
where $W_{i}$ is given by (\ref{DefW}). $X$ can be written as
\be \label{Xnew}
X=|I\ra\la \phi^* \phi^*|,
\ee
which implies that
\be
EX=X.
\ee
The MPS representation allows a simple and efficient calculation of local observables and correlation functions \cite{MPS}. This is done by assigning to any local operator $A=\sum_{i,j}A_{ij}|i\ra\la j|$ operators in the auxiliary vectorized space
\be \label{DefEA1}
E_{A}=\sum_{i,j} \langle i|A|j\rangle V_{i}^{*} \otimes V_{j},
\ee
and
\be \label{DefEA2}
X_{A}=\sum_{i,j} \langle i|A|j\rangle W_{i}^{*} \otimes W_{j}.
\ee
and then expressing any one-point function in the following form
\ba\label{One-point1}
\langle A_{m} \rangle &=&tr(E^{m-1}E_{A}E^{N-m-1}X) \nonumber \\
&=&tr(E^{m-1}E_{A}X), \hskip 0.5cm m\neq N,
\ea
and
\be \label{One-point2}
\langle A_{N}\rangle=tr(E^{N-1}X_{A}),
\ee
where in (\ref{One-point1}) we have used the fact that $EX=X$. Two-point functions can be calculated in the same way with the result (using the fact that $EX=X$)
\begin{eqnarray}\label{Two-point}\nonumber
\langle A_{m} A_{n} \rangle=tr(E^{m-1}E_{A}E^{n-m-1}E_{A}X), \hskip 0.1cm m,n \neq N,\\
\langle  A_{m} A_{N} \rangle=tr(E^{m-1}E_{A}E^{N-m-1}X_{A}), \hskip 0.7cm m\neq N.
\end{eqnarray}
\section{Measure of macroscopicity for the sequentially generated state}\label{Sec4}
We are now in a position to use the power and elegance of the matrix product formalism to calculate the measure of macroscopicity of the state (\ref{psii}) and find an expression for its effective size. To do this we start from the variance of an additive operator $A=\sum_{m}A_{m}$ (with $A_m^2=1$) and write
\begin{eqnarray} \label{variance}
(\bigtriangleup A )^{2}=\sum_{m,n=1}^{N}(\langle A_{m}A_{n} \rangle -\langle A_{m}\rangle \langle A_{n} \rangle).
\end{eqnarray}
The effective size is determined by (\ref{effSize}) or
\begin{equation}\label{}
N_{\text{eff}}(\psi)=\max_{A} \frac{\sum_{m,n=1}^{N}(\langle A_{m}A_{n} \rangle -\langle A_{m}\rangle \langle A_{n} \rangle)}{N}.
\end{equation}
A state is a macroscopic superposition if its effective size is proportional to the size $N$, hence to assess the macroscopicity of this state, we can ignore the terms linear and sublinear in the numerator such as two point functions where one of the points is in the bulk and the other is in the boundary (\ref{Two-point}). Thus we can write $(\bigtriangleup A )^{2}$ as follows
\ba \label{varianceMPS}
(\bigtriangleup A )^{2}&=&2 \sum_{1\leq m<n<N}tr(E^{m-1}E_A E^{n-m-1}E_AX) \nonumber \\
&-&(\sum_{1\leq m<N}tr(E^{m-1}E_AX))^2+{\cal O}(N).
\ea
As discussed in the previous section, the eigenvalues of the linear operator $E$ are either unity, or their absolute value is strictly less than unity. For (\ref{varianceMPS}) to yield a dependence that is quadratic in $N$, the powers $E^{m-1}, E^{n-m-1}$ must hence be restricted to their action on the unit eigenvalue eigenspace, i.e.,  $E$ and all high powers of the same operator may be simply replaced by the projection on this space. The quadratic in $N$ dependence thus obtains $\frac{(N-1)(N-2)}{2}$ and $(N-1)^2$ identical contributions from the first and  second term in  (\ref{varianceMPS}), respectively.

\textbf{Remark:} Anticipating a possible degeneracy of the unit eigenvalues of $E$ we denote its eigenvectors by $|{\bf 0}\ra \propto |I\ra(= |00\ra+|11\ra)$ and $|\tilde{\bf{0}}\ra$.

If the unit eigenvalue is non-degenerate, the right eigenvector $|{\bf 0}\ra \propto |I\ra(= |00\ra+|11\ra)$, while the left eigenvector $\la {\bf 0}|$ depends on the unitary operator $U$, and we have
\be
E^k \approx |{\bf 0}\ra\la {\bf 0}|, \textrm{for large $k$}.
\ee
Assuming the normalization $\la {\bf 0}|{\bf 0}\ra=1$, we find
\be
N_{\text{eff}}(\psi)=\max_{A}\left[\la {\bf 0}|E_A|{\bf 0}\ra \la {\bf 0}|E_AX|{\bf 0}\ra-\la {\bf 0}|E_AX|{\bf 0}\ra^2\right]\ N,
\ee
 Since according to (\ref{XMatrix}) $X|{\bf 0}\ra=|{\bf 0}\ra$,  the numerical pre-factor vanishes and no macroscopicity is produced by the sequential operation of $U$ on the chain.

However, when the unit eigenvalue is degenerate with two right eigenvectors $|{\bf 0}\ra$ and $|\tilde{\bf{0}}\ra$, we have
\be
E^k \approx |{\bf 0}\ra\la {\bf 0}|+ |\tilde{\bf{0}}\ra\la \tilde{\bf{0}}|, \hskip 0.5cm \textrm{for large $k$}.
\ee
Now, we get
\ba \label{Neff}
N_{\text{eff}}(\psi)=\max_{A}[\la {\bf 0}|E_A|\tilde{\bf{0}}\ra \la \tilde{\bf{0}}|E_A|{\bf 0}\ra+\la \tilde{\bf{0}}E_A|{\bf 0}\ra \la {\bf 0}|E_AX|\tilde{\bf{0}}\rangle \nonumber \\ 
+\la \tilde{\bf{0}}|E_AX|\tilde{\bf{0}}\ra (\la \tilde{\bf{0}}|E_A|\tilde{\bf{0}}\ra-2\la {\bf 0}|E_A|{\bf 0}\ra-\la \tilde{\bf{0}}|E_AX|\tilde{\bf{0}}\ra)]N,
\ea
and we note that in view of (\ref{Xnew}) and the remark above, $X|\tilde{\bf{0}}\ra \propto |{\bf 0}\ra$, so we obtain a non-zero effective size. We shall now study a few explicit examples.


\subsection{Results for a class of symmetric unitary nearest neighbor operations}
We address the macroscopicity (\ref{Neff}), produced by sequential application of a two-qubit unitary on an initial product state, $|0\rangle^{N}$. Delegating the detailed calculation to appendix (\ref{appendix a}) we consider unitaries of the form \cite{Whaley}
\begin{equation} \label{weyl}
U_{n,n+1}=e^{-\frac{i}{2}(\alpha \sigma_n^{x} \otimes \sigma_{n+1}^{x}+\beta \sigma_n^{y} \otimes \sigma_{n+1}^{y}+\gamma \sigma_n^{z} \otimes \sigma_{n+1}^{z})}.
\end{equation}
In appendix A we obtain the eigenvalues of $E$: $1$, $\sin \alpha \sin \beta$, $\frac{1}{2}\sin\gamma(\sin \alpha+\sin \beta)\pm \frac{1}{2}\sqrt{\sin\gamma^{2}(\sin \alpha+\sin \beta)^{2}-4\sin\alpha\sin\beta}$. These eigenvalues are bounded by unity as we concluded above, and to obtain a macroscopic state, Eq. (\ref{Neff}), we require that at least two eigenvalues equal unity. This occurs for example if $\beta=\gamma=\frac{\pi}{2}$, and we obtain
\begin{equation}\label{}
N_{\text{eff}}\equiv \frac{(\bigtriangleup A_{\bf y})^{2}}{N}=(\cos^{2}\alpha) N,
\end{equation}
where $A_{\bf y} = \sum_{n=1}^N \sigma^y_n$.

If we assume instead $\alpha=\gamma=\frac{\pi}{2}$, we get
\begin{equation}\label{}
N_{\text{eff}}\equiv \frac{(\bigtriangleup A_{\bf x})^{2}}{N}=(\cos^{2}\beta) N,
\end{equation}
where $A_{\bf x} = \sum_{n=1}^N \sigma^x_n$.

\subsection{Result for controlled-unitary nearest neighbor operations}

The fluctuations of a collective observable may be large, but increases linearly with the number of particles $N$, if each particle is correlated with a finite number of other particles. We can illustrate the role of the correlation over longer distances with the particular example of controlled unitary operations, acting on the $n+1^{st}$ particle conditioned on the $n^{th}$ particle being in state $|1\rangle$.
Starting from our initial product state with $|\phi\rangle = \frac{1}{\sqrt{2}}(|0\ra + |1\ra)$, and all other qubits being in state $|0\ra$,  see Fig.2, and applying the controlled unitary $U=|0\rangle_i\langle 0|\otimes \mathbbm{1}_{i+1}+ |1\rangle_i\langle 1|\otimes U^0_{i+1}$ with
\begin{equation}\label{DefU0}
U^{0}=\left(
        \begin{array}{cc}
          \cos \frac{a}{2} & -\sin \frac{a}{2} \\
          \sin \frac{a}{2} & \cos \frac{a}{2} \\
        \end{array}
      \right).
         \end{equation}

The gate executes a perfect C-NOT operation, and a perfectly correlated GHZ state of the qubits is created, if $\alpha=\pi$. For smaller or larger values of $\alpha$ the conditional rotation does not cause a complete switch of the target qubit state into $|1\rangle$ and, hence, the probability that $U_0$ is applied to the subsequent qubits will generally decrease along the chain. If the length of the chain is short, collective observables may still show fluctuations quadratic in $N$, but when the chain gets longer, we observe a passage to a constant dependence. This is illustrated in Fig.3, where the collective  variance of $A_z$ is shows as a function of $N$ for different single qubit rotation angles $\alpha$.

\begin{figure}[h] \label{circuit}
\centering
\includegraphics[width=5.5cm,height=3.5cm,angle=0]{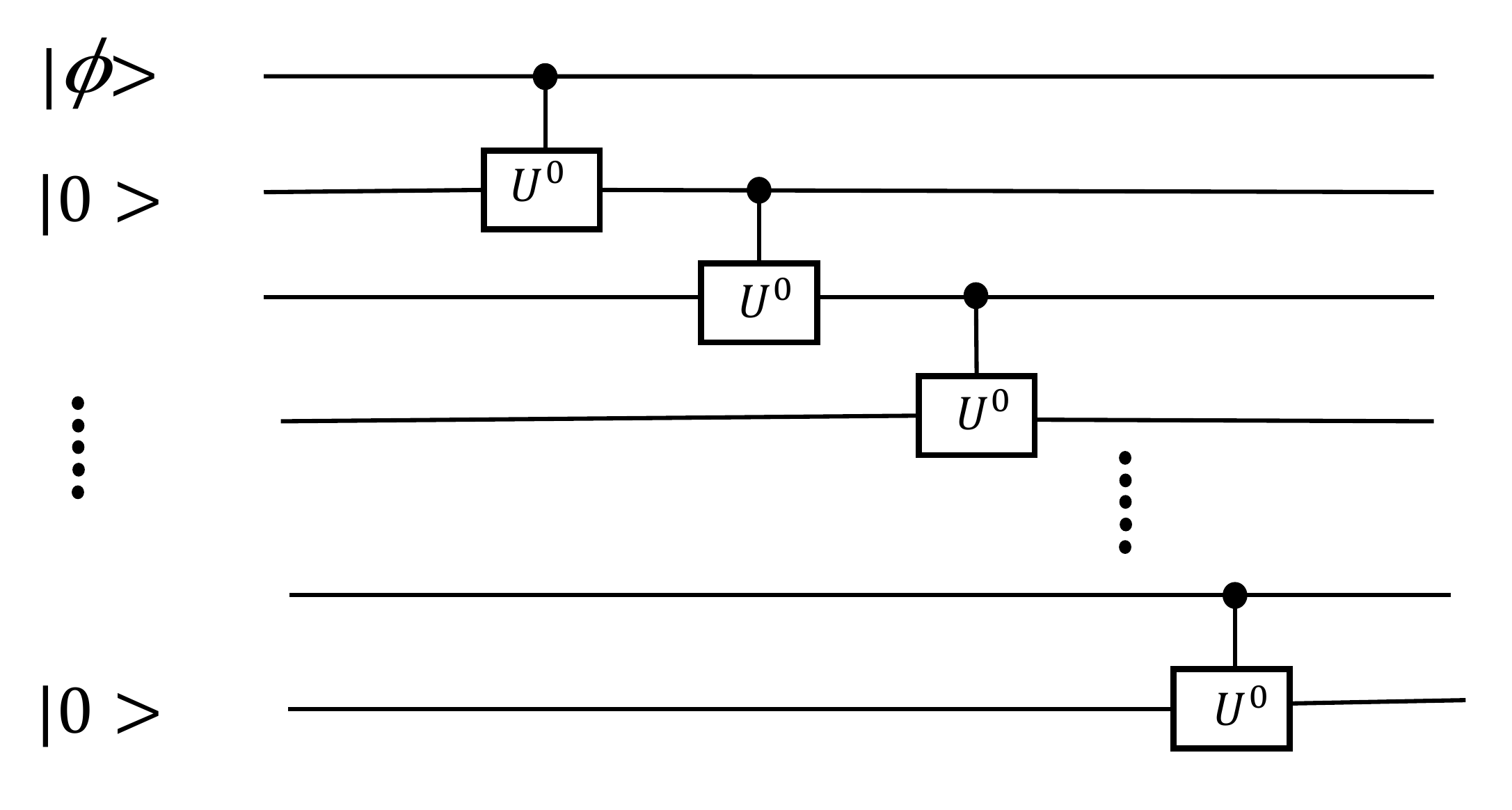}
\caption{The controlled-$U^{0}$ operators, acting on the target qubit if and only if the control qubit is $|1\rangle$, are sequentially applied between nearest neighbour qubits. We choose the first qubit in $|\phi\rangle=\frac{|0\rangle+|1\rangle}{\sqrt{2}}$.}
\end{figure}

\begin{figure}[h]\label{CNOT}
\centering
\includegraphics[width=5.5cm,height=4.5cm,angle=0]{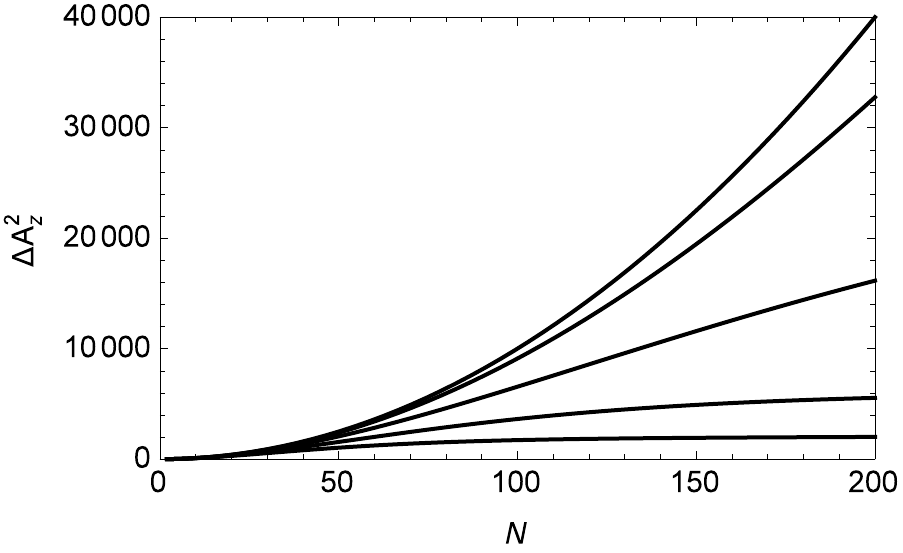}
\caption{The collective  variance of $A_z$ as a function of $N$. The curves from top to bottom, correspond to rotation by $a= \pi, \pi-0.1, \pi-0.2, \pi-0.3, \pi-0.4$. For all angles, we observe a quadratic behavior for small values of $N$, but as we gradually increase the value of $N$, the scaling becomes linear except in the $\pi$-rotation which leads to controlled-NOT operation.}
\end{figure}

\subsection{General form of unitary gates leading to macroscopicity}

Given a unitary operation $U$, we can use (\ref{DefV}) and (\ref{DefE1}) to evaluate the eigenvalues of $E$ and determine if the operation leads to macroscopic superposition state. In this section we want to identify the general form of the unitary two qubit gates which generate macroscopic superposition i.e., that lead to the doubly degenerate largest eigenvalue of $E$. In view of the correspondence (\ref{unitalMap}) with unital maps, we first start from the following theorem:

{\bf Theorem:}

The only unital qubit channels (with two Kraus operators) having degenerate unit eigenvalues are of the form
\be
{\cal E}(\sigma)=(1-p)\bm \omega \sigma \bm\omega^{\dagger} + p \bm\omega' \sigma \bm\omega'^{\dagger},
\ee
where $\bm\omega$ and $\bm\omega'$ are two commuting single-qubit unitary operators, i.e. rotations around the same axis with different angles.

{\bf Remark:} More generally such channels can be of the form $\sum_i p_i \omega_i \sigma \omega_i^\dagger$, where all $\omega_i$'s commute with each other. We prove the theorem for the simplest case of two Kraus operators, the general proof is similar.

{\bf Proof:} Let ${\cal E}$ be a channel of the form
\be\label{EEE}
{\cal E}(\sigma)=V_0\sigma V_0^\dagger + V_1 \sigma V_1^\dagger,
\ee
which has two different eigenvector with unit eigenvalue, that is
\be\label{myeig}
{\cal E}(I)=I,\h {\cal E}(Y)=Y,
\ee
where the first relation is the definition of unitality and $Y$ is not necessarily a density matrix.

First we note that if $Y$ is an eigenvector, then $Y^\dagger$ is also an eigenvector with the same eigenvalue. Hence any unit eigenvector can be taken to be Hermitian and hence of the form $\lambda I+ {\bf r}\cdot \bm \sigma$, where $\lambda$ is a real number and ${\bf r}$ is a real vector. Since  the channel is unital we can make a suitable combination of $I$ and this eigenvector and normalize it to a unit eigenvector of the form $\frac{1}{2}(I+{\bf n}\cdot\bm\sigma)$, where ${\bf n}$ is a unit vector. But this is nothing but the pure state $|{\bf n}\ra\la {\bf n}|$, where $|{\bf n}\ra$ is the positive spin state in the ${\bf n}$ direction. Therefore from (\ref{myeig}) it follows that
\be
V_0|{\bf n}\ra\la {\bf n}|V_0^\dagger + V_1 |{\bf n}\ra\la {\bf n}| V_1^\dagger=|{\bf n}\ra\la {\bf n}|,
\ee
that is a pure state is written as a convex combination of two other states. However this is only possible if the two other states are multiples of the same pure state. This happens only if both  matrices $V_0$ and $V_1$ leave $|{\bf n}\ra$ invariant, that is ${\bm \omega}=e^{i\theta {\bf n}\cdot \bm\sigma}$ and ${\bm \omega}'=e^{i\theta' {\bf n}\cdot \bm\sigma}$. This completes the proof.

We can now determine the explicit form of the two-qubit unitary gates which generate macroscopic superposition. To this end use the relation (\ref{DefV}) and the explicit form of the two rotations
\be 
\bm \omega = \left(\begin{array}{cc}\cos\theta & i\sin\theta \\  i\sin \theta & \cos \theta\end{array}\right), \hskip 0.4cm
 \bm \omega = \left(\begin{array}{cc}\cos\theta' & i\sin\theta' \\ i\sin \theta' & \cos \theta' \end{array}\right), \nonumber
\ee
 to write $U$ as
\be \label{U}
U=\left(\begin{array}{cccc}(1-p)\cos \theta & . & i(1-p)\sin\theta & . \\
i(1-p)\sin \theta & . & (1-p)\cos\theta & . \\
p\cos \theta' & . & ip\sin\theta' & . \\
ip\sin \theta' & . & \cos\theta' & .
\end{array}\right),
\ee
where the second and fourth column entries can be chosen freely, subject to the unitarity of the two qubit gate $U$.
Any unitary of the form
\begin{equation}\label{}
{\cal U}=(R_{1}\otimes R_{2})U(R_{1}^{\dagger}\otimes R_{2}^{\dagger}),
\end{equation}
where $U$ is given by (\ref{U}), and $R_1$ and $R_2$ are single qubit rotations, lead to macroscopicity.

\section{Squeezing of collective spin by sequential nearest neighbour interactions}\label{Sec5}

Kitagawa and Ueda \cite{Kitagawa} proposed to produce spin squeezed states by subjecting a large spin to a “two-axis twisting” Hamiltonian, $H \propto (J_{x}^{2}-J_{x}^{2})$, implemented in a collection of spin 1/2 particles where each spin interacts in the same way with all other spins, e.g., $J_x^2 = \sum_{ij} \frac{1}{4} \sigma^x_i\sigma^x_j$. Motivated by \cite{Klaus}, which shows that a linear chain of spin 1/2 particles also becomes squeezed if subject only to a nearest neighbor interaction $H=\sum_i \frac{\chi}{2}(\sigma_{i}^{x}\sigma_{i+1}^{x}-\sigma_{i}^{y}\sigma_{i+1}^{y})$, 
we shall study the sequential application of unitary operations $U=e^{-iHt}$, where
\be \label{H}
H=\frac{\chi}{2}(\sigma_{i}^{x}\sigma_{i+1}^{x}-\sigma_{i}^{y}\sigma_{i+1}^{y}).
\ee

We use the MPS formalism to investigate if the sequential application of the pairwise interaction Hamiltonian also leads to squeezing.
The matrix $E$, and its eigenvalues, $1>\sin \chi t>-\sin^{2} \chi t>-\sin \chi t$ and eigenvectors are determined in Appendix B. 
We further provide the matrix $E_A$, with $A=J_{\bf{n}} = \bf{n}\cdot \bf{J}$ representing the component of the collective spin along an arbitrary unit vector ${\bf n}$.
Starting in the state $|0\ra^N$, we readily find (and it also follows from Heisenberg's equations of motion) that the mean values of the $x$- and $y$-components of the individual and collective spin vanish identically for all times. For large $N$, the mean value of $J_z$ is dominated by the term proportional with $N$
\be \label{SzSqueezing}
\langle \psi|J_z|\psi \rangle\simeq \la{\bf 0}|E_{J_z}|{\bf 0 }\ra N= \frac{1-3\sin^2(\chi t)}{1+ \sin^2 (\chi t)}N.
\ee
We consider now the variance of the collective spin observables, orthogonal to $J_z$ (4). For a large number of spin qubits, we recover a term linear in $N$ from the single site variances (terms with $m=n$), while the more complicated two-site correlations demand a careful treatment.
\ba
(\bigtriangleup J_{\theta})^{2}&=&N+2\sum_{1\leq m<n<N}tr(E^{m-1}E_{J_{\theta}} E^{n-m-1}E_{J_{\theta}}X) \nonumber \\ 
&+&{\cal O}(1),
\ea

Unlike our analysis of macroscopicity, where only the unit eigenvalues of $E$ mattered, the lower power in $N$ gets contributions from all eigenvalues $\lambda_{i}$ ,$i=1,2,3,4$ , and it is convenient to expand the expressions in terms of the associated eigenvectors of $E$, $\{|\lambda_{i}\rangle\}$, 
\ba \label{sumapp}
(\bigtriangleup J_{\theta})^{2}&=&2\sum_{i,j=1}^{4}\sum_{n=2}^{N-1}\sum_{m=1}^{n-1} (E_{J_{\theta}})_{ij}(E_{J_{\theta}}X)_{ji}\lambda_{i}^{m-1}\lambda_{j}^{n-m-1}\nonumber \\ 
&+&N+{\cal O}(1).
\ea
The summations over sites and eigenvalues is carried out in Appendix B, and shows that the spin component $J_{\frac{\pi}{4}}=\frac{1}{\sqrt{2}}(J_{x}+J_{y})$ is squeezed, and we get
\be \label{VarSqueezing}
(\bigtriangleup J_{\frac{\pi}{4}})^{2}=[1-\frac{2\sin(2\chi t)\cos(\chi t)}{(1+\sin^2(\chi t))(1+\sin(\chi t))}]N+{\cal O}(1).
\ee
The ratio between $(\bigtriangleup J_{\frac{\pi}{4}} )^{2}$ and $\la J_z \ra$ witnesses, through (4) the squeezing and the entanglement in the system, but plotting these values normalized by the number of spins $N$ as $x$ and $y$ coordinates for different values of the accumulated interaction $\chi t$ leads to the dashed red curve in Fig.4, which characterizes further properties of the entanglement. The solid black line in the figure shows the minimum variance of $J_\theta$ given the mean value of $J_z$ of any separable state of the spins, while the dotted blue curve in the figure shows the minimum achievable variance of $J_{\theta}$ given the value of $J_{z}$ (both normalized by $N$) for an ensemble of spin 1/2 particles that may be entangled but only in pairs \cite{Sorensen}. The fact that the dashed red curve lies below the dotted blue one in the figure for large values of $\la J_z\ra$,  witnesses the presence of multi-partite entanglement, the spins in the chain must at form groups of least three entangled spins. 

\begin{figure}[h]\label{entanglementDepth}
\centering
\includegraphics[width=5.5cm,height=4.5cm,angle=0]{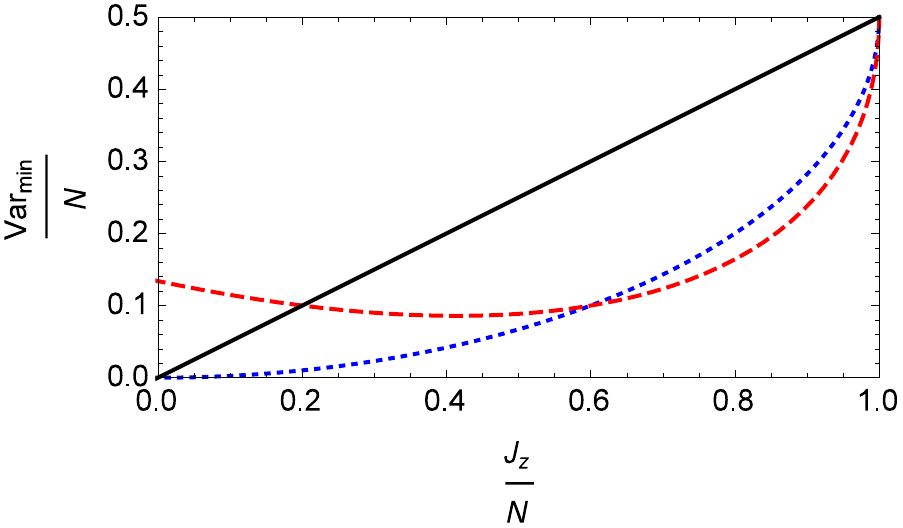}
\caption{Maximal collective squeezing in the limit of large $N$. The dashed red curve represents accompanying values of the mean spin $\la J_z \ra$ and the minimum transverse spin variance $(\bigtriangleup J_{\frac{\pi}{4}} )^{2}$, while the solid black(dotted blue) curve shows the minimum transverse variance $(\bigtriangleup J_{\theta} )^{2}$ allowed for given $\la J_z \ra$ if the spins are separable (at most pairwise entangled) \cite{Sorensen}. For large $J_{z}$ the dashed red curve lies below the dotted blue one, so the spin squeezed state generated by sequential pairtwise interaction contains at least three-spin entangled components.}
\end{figure}

\section{Conclusion}\label{Sec6}

We have used the matrix product state description to calculate local mean values and two-site correlations function of observables for a chain of particles prepared by a sequential nearest neighbor interactions. After a single sweep of such interactions acting on an initial product state, the chain can be explicitly described by a matrix product state with the same matrix dimension as the Hilbert space of the individual particles. We used the results to determine the behavior of collective properties such as macroscopicity, collective spin squeezing and multiparticle entanglement and we derived criteria on the pairwise interaction parameters for the observation of these properties. While we focussed on two-level systems in our examples, the formalism is general and allows treatment of general $d$-level systems by similar expressions. By grouping the systems in pairs, nearest neighbor interactions between these larger systems can represent any next nearest neighbor coupling of the original quantum systems, which can thus also be studied by our formalism. 

There is a formal connection between the Matrix Product State description of
one-dimensional chain systems and the time evolution of the density matrix of a single quantum system under a trace preserving, completely positive map. This correspondence is particularly strong in the case of sequential operations, and, indeed, the trace preserving  map, defined in Eq.(\ref{dualmap}), recursively provides the reduced density matrix $\rho_{n}$ of the $n$th particle after the action of $U_{n-1,n}$ in terms of the previous, reduced density matrix $\rho_{n-1}$ of the $n-1^{st}$ particle after the first $n-1$ particles have been subject to the interactions.
This follows from the explicit form
\be
\rho_{n}=tr_{n-1}(U (\rho_{n-1} \otimes (|0\rangle \langle 0|)_n)U^{\dagger}).
\ee
In the same way as a time evolving density matrix permits evaluation of time dependent expectation values, our map permits evaluation of the site dependent expectation values. And, in the same was as the Quantum Regression Theorem applies the propagator for the density matrix to evaluate temporal correlation functions, we are able to compute spatial correlation functions along the chain.\\
A final interesting connection between time evolving density matrices and the spatially growing MPS, concerns our criterion of a degenerate unit eigenvalue of the unital map ${\cal E}$ to observe macroscopicity. The resulting degeneracy of the unit eigenvalue of ${\cal E}^*$ ensures that the qubits along the chain do not converge to a single density matrix, i.e., the system may retain its correlation with the state of the first qubit indefinitely along the chain. A similar property of density matrix evolution in the time domain implies that the quantum system may not have a definite steady state, and hence it may have an infinitely long memory of its earlier states. The degeneracy of the unit eigenvalue of the propagator of the master equation has, indeed, been identified as the source of a Fisher information that scales as $T^2$ rather than $T$ for continuous probing of a system for a time $T$ \cite{Madalin, Gammelmark, Macieszczak} and hence the possibility to probe system parameters with a variance scaling as $1/T^2$ rather than $1/T$.

\subsection*{Acknowledgements} T.A. and K.M. acknowledge support from the Villum Foundation, and T.A. acknowledges support from the Ministry of Science, Research and Technology of Iran and Iran Science Elites Federation.

{}

\appendix
\section{Details of calculation for section \ref{Sec4}}\label{appendix a}
The unitary operator in Eq.(\ref{weyl}) has the matrix representation 
\begin{equation} \label{u}
U=\left(
        \begin{array}{cccc}
          x & 0 &0&w\\
          0 & z &y&0\\
          0 &y&z&0\\
        w&0&0&x
        \end{array}
      \right),
         \end{equation}
where
\ba \label{xyzw}
x=e^{-\frac{i \gamma}{2}}\cos (\frac{\alpha-\beta}{2}), \hskip 0.7cm y=-i e^{\frac{i \gamma}{2}}\sin (\frac{\alpha+\beta}{2}), \nonumber \\ 
z=e^{\frac{i \gamma}{2}}\cos (\frac{\alpha+\beta}{2}), \hskip 0.7cm  w=-i e^{-\frac{i \gamma}{2}}\sin (\frac{\alpha-\beta}{2}).
\ea
We constitute matrices $V_0$ and $V_1$ from  Eq. (\ref{DefV}),
\begin{equation} \label{V0V1}
V_{0}=\left(
        \begin{array}{cc}
          x & 0 \\
          0 & y\\
        \end{array}
      \right),  \hskip 1cm V_{1}=\left(
             \begin{array}{cc}
               0 & w \\
              z & 0 \\
             \end{array}
           \right).
         \end{equation}
Using (\ref{DefE1}) it is straightforward to find the matrix $E$,
\begin{equation} \label{EE}
E=\left(
        \begin{array}{cccc}
          |x|^2 & 0 &0&|w|^2\\
          0 & x^*y &zw^* &0\\
          0 & z^*w &  xy^*  &0\\
        |z|^2&0&0&|y|^2
        \end{array}
      \right),
         \end{equation}
and determine its eigenvalues: $1,\ \sin \alpha \sin \beta,\ \frac{1}{2}\sin\gamma(\sin \alpha+\sin \beta)\pm \frac{1}{2}\sqrt{\sin\gamma^{2}(\sin \alpha+\sin \beta)^{2}-4\sin\alpha\sin\beta}$ To obtain two degenerate eigenvalues, $1$, we have to set $\beta=\gamma=\frac{\pi}{2}$ or $\alpha=\gamma=\frac{\pi}{2}$, with two eigenvectors,
\ba \label{}\nonumber 
|\bf{0}\ra&=&(|00\ra+|11\ra), \hskip 0.8cm \la \textbf{0}|=\frac{1}{2}(\la 00|+\la 11|)  \\
|\tilde{\bf{0}}\ra&=&(\pm|01\ra+|10\ra), \hskip 0.55cm \la \tilde{\bf{0}}|=\frac{1}{2}(\pm \la 01|+\la 10|),\nonumber
\ea
and from (\ref{Neff}) we obtain,
\begin{equation}\label{N}
N_{\text{eff}}=\max_{A}| \la \tilde{\bf{0}}|E_A|{\bf 0}\ra|^2N,
\end{equation}
where we use $X|\tilde{\bf{0}}\ra=0$. To proceed further, using the fact that, (\ref{V0V1}), $V_0|i\ra\propto |i\ra$ and $V_1|i\ra\propto|1-i\ra$ for $i=0,1$, we find from (\ref{DefEA1}) that
\ba \nonumber
E_{A}|00\ra&=&x^*zA_{01}|01\ra+xz^*A_{10}|10\ra +|a \ra,\\
E_{A}|11\ra&=&y^*wA_{01}|10\ra+yw^*A_{10}|01\ra +|b \ra.\nonumber
\ea
where $A_{01}=\la 0|{\bf n}\cdot \bm{\sigma}|1\ra$, and $|a \ra, |b \ra$ are linear combination of $|00\ra$ and $|11\ra$. Finally we get from (\ref{N}),
\begin{equation}\label{} \nonumber
N_{\text{eff}}=\frac{1}{4}\max_{A}|\pm A_{01}(x^*z+y^*w)+A_{10}(xz^*+yw^*)|^2N.
\end{equation}
Eq. (\ref{xyzw}) and $\gamma=\frac{\pi}{2}$ lead to
\ba \nonumber
N_{\text{eff}}=\frac{1}{4}\max_{A}|(- A_{01}+A_{10})\cos \a|^2N, \hskip 0.4cm \beta=\frac{\pi}{2},\\
N_{\text{eff}}=\frac{1}{4}\max_{A}|( A_{01}+A_{10})\cos \b|^2N, \hskip 0.7cm \alpha=\frac{\pi}{2}.\nonumber
\ea
So we get 
\ba \nonumber
N_{\text{eff}}&=&\frac{(\bigtriangleup A_{\bf y})^{2}}{N}=(\cos^{2}\alpha) N, \hskip 0.4cm \gamma=\beta=\frac{\pi}{2},\\
N_{\text{eff}}&=& \frac{(\bigtriangleup A_{\bf x})^{2}}{N}=(\cos^{2}\beta) N, \hskip 0.4cm \gamma=\alpha=\frac{\pi}{2},\nonumber
\ea
where $A_{\bf y} = \sum_{n=1}^N \sigma^y_n$ and $A_{\bf x} = \sum_{n=1}^N \sigma^x_n$.


\section{Details of calculation for section \ref{Sec5}}\label{appendix b}
The unitary operations $U=e^{-iHt}$, (\ref{H}),  is a special case of Eq. (\ref{u}) with $(x=\cos \chi \tau, y=0, z=1, w=-i\sin \chi \tau)$, and with diagonalization of $E$, (\ref{EE}), we get  the eigenvalues: $1, iw, w^2, -iw$, with right and left eigenvectors,
\begin{eqnarray}\label{eig}\nonumber
|{\bf 0}\rangle=|00\rangle+|11\rangle,  \hskip 0.3cm \la{\bf 0}|=\frac{1}{1+|w|^2}(\la00|+|w|^2\la11|),\\ \nonumber
|\tilde{\bf{0}}\rangle=i|01\rangle+|10\rangle, \hskip 1.5cm \la\tilde{\bf{0}}|=\frac{1}{2}(-i\la01|+\la10|),\\ \nonumber
|{\bf 2}\rangle=-i|01\rangle+|10\rangle, \hskip 1.5cm \la{\bf 2}|=\frac{1}{2}(i\la01|+\la10|),\\ 
|{\bf 3}\rangle=\frac{1}{1+|w|^2}(-|w|^2|00\rangle+|11\rangle),  \hskip 0.1cm \la{\bf 3}|=-\la00|+\la11|.
\end{eqnarray}
To find the mean value of $J_z$, first we find matrix $E _{\textbf{J}}$ using (\ref{DefEA1}),
\be \label{EJ} \nonumber
E _{\textbf{J}}=\left(\begin{array}{cccc}n_z |x|^2 & ( n_x-i n_y)x^*w &( n_x+i n_y)xw^* & -n_z |w|^2 \\ 
(n_x-i n_y)x^* & 0 & -n_z w^* & 0 \\
(n_x+i n_y)x & -n_z w &0 & 0 \\
-n_z & 0 & 0 & 0 
\end{array}\right).
\ee
So we get
\be \label{} \nonumber
\langle \psi|J_z|\psi \rangle\simeq \la{\bf 0}|E_{J_z}|{\bf 0 }\ra N= (\frac{|x|^2-2|w|^2}{1+|w|^2})N.
\ee 
which leads to (\ref{SzSqueezing}),
\be \label{} \nonumber
\langle \psi|J_z|\psi \rangle\simeq\frac{1-3\sin^2(\chi t)}{1+ \sin^2 (\chi t)}N.
\ee
To find $(\bigtriangleup J_{\theta})^{2}$, the summations over sites in (\ref{sumapp}) turns out to be
\be \label{var} \nonumber
(\bigtriangleup J_{\theta} )^{2}=N+2\sum_{i,j=1}^{4} (E_{J_{\theta}})_{ij}(E_{J_{\theta}}X)_{ji}f(\lambda_{i,j}, N)+{\cal O}(1),
\ee
where
\begin{equation}\label{f} \nonumber
f(\lambda_{i,j}, N)=\frac{(\lambda_{i}-\lambda_{j})-\lambda_{i}^{N-1}(1-\lambda_{j})+(1-\lambda_{i})\lambda_{j}^{N-1}}{(\lambda_{i}-\lambda_{j})(1-\lambda_{i})(1-\lambda_{j})},
\end{equation}
where the scaling of $f(\lambda_{j}, N):=f(\lambda_{i}=1,\lambda_{j}, N)$ is found to be
\begin{equation}\label{} \nonumber
f(\lambda_{j}, N)=\frac{N}{1-\l_{j}}+{\cal O}(1), \hskip 1cm  j\neq 1.
\end{equation}
So $(\bigtriangleup J_{\theta})^{2}$ attains a term linear in $N$,
\be  \label{KL}
(\bigtriangleup J_{\theta})^{2}=[1+2\sum_{j=2}^{4}\frac{(E_{ J_{\theta}})_{1j}(E_{ J_{\theta}})_{j1}}{1-\l_{j}}]N+{\cal O}(1),
\ee
where we use $X|{\bf 0}\ra=|{\bf 0}\ra$. To proceed further we note that for the transverse component $J_\theta = n_x J_x + n_y J_y$, we have
\be \nonumber
E_{ J_{\theta}}|{\bf 0}\ra=(n_x-i n_y)x^* |01\ra+(n_x+i n_y)x |10\ra,
\ee
and
\be \nonumber
\la {\bf 0}|E_{ J_{\theta}}=(n_x-i n_y)x^* w\la01|+(n_x+i n_y)xw^* \la10|.
\ee
So using (\ref{eig}) and (\ref{KL}) we obtain
\ba  \label{VV}
(\bigtriangleup J_{\theta})^{2}&=&[1+(\frac{2x^2(iw)}{1+|w|^2})(\frac{(n_x-n_y)^2}{1-iw}-\frac{(n_x+n_y)^2}{1+iw})]N \nonumber \\
&+&{\cal O}(1),
\ea
where we use the fact that $x$ and $w$ are real and imaginary numbers, respectively. To minimize (\ref{VV}), we have to set $n_x=n_y=\frac{1}{\sqrt{2}}$, and we get
\be \nonumber
(\bigtriangleup J_{\theta})^{2}=[1-\frac{4x^{2}(iw)}{(1+|w|^2)(1+iw)}]N+{\cal O}(1),
\ee
which leads to (\ref{VarSqueezing}),
\be \nonumber
(\bigtriangleup J_{\theta})^{2}=[1-\frac{2\sin(2\chi t)\cos(\chi t)}{(1+\sin^2(\chi t))(1+\sin(\chi t))}]N+{\cal O}(1).
\ee

\end{document}